\documentclass[doublecol]{epl2} 

\title{Minimalistic analytical approach to non-Markovian open quantum systems}
\shorttitle{Effective study of non-Markovian  open quantum systems} 

\author{Vitalii Semin\inst{1,2} \and Francesco Petruccione\inst{1,2}}
\shortauthor{V. Semin and F. Petruccione}

\institute{                    
  \inst{1} Quantum Research Group, School of Chemistry and Physics,
 University of KwaZulu-Natal, Durban, 4001, South Africa\\
  \inst{2} National Institute for Theoretical Physics (NITheP), KwaZulu-Natal, South Africa
}
\pacs{03.65.Yz}{Decoherence; open systems; quantum statistical methods}

\abstract{The dynamics of {finite dimension} open quantum systems is studied with the help of the simplest possible form of projection operators, namely the ones which project only onto one dimensional subspaces. The simplicity of the action of the projection operators always leads to an analytical solution of the dynamical master equation, even in the non-Markovian case, in any perturbative order. The analytical solution correctly reproduces the short-time dynamics, and can be used to recursively recover the dynamics for an arbitrary time interval {with arbitrary precision}. The necessary number of relevant degrees of freedom to completely characterise an open quantum system is $(n-1)(n+2)/2,$ where $n$ is  the dimension of the Hilbert space of the open system.  The method is illustrated by two examples, the relaxation of a qubit in a thermal bath and the dynamics of two interacting qubits in a common environment.}

\begin{document}

\maketitle

\section {Introduction}
The understanding of  the dynamics of open quantum systems is necessary to describe  many interesting phenomena such as photosynthesis \cite{PS},  the transport in living cells \cite{transfer} and the dynamics of quantum systems in strong laser fields \cite{LF}. Recently, several different approaches to study open systems  have been suggested \cite{MY1, KOSSAK, KOCH, toqs, CAO, BUDINI, BREUER}. These approaches significantly differ from each other and describe certain properties of open systems from different points of view. By choosing the most appropriate method to describe an open system one may successfully study the relevant characteristics of the system with a reasonable accuracy. Despite the significant success of the theoretical investigation it is still difficult to derive analytical or, even, numerical results  for a general open system, especially, in the non-Markovian case.  Typically, non-Markovian master equations have very complicated dependence on time and even a numerical study of such equations is a non-trivial task.

In this letter we suggest an approach, based on a special class of projection operators, which allows to study  the dynamics of a broad class of open systems. Application of the suggested technique to  {finite dimension} open quantum systems always leads to an integrable set of differential equations. The number of the equations, which are necessary to characterise an open system, is lesser than the dimension of the reduced density operator of the open system. The analytical solution correctly reproduces the short-time dynamics of an open quantum system, and under additional assumptions allows to recursively recover the solution of traditional forms of master equations {with arbitrary precision}. The method is illustrated by two examples, relaxation of a qubit in a thermal bath and relaxation of two interacting qubits in a common environment.

\section{Time-convolutionless master equation}
Our approach is based on the so-called time-convolutionless (TCL) master equation. The most widely used form of the TCL master equation reads \cite{toqs}
\begin{equation}\label{TCL}
\mathcal{P}\dot{\rho}=\mathcal{K}(t)\mathcal{P}\rho(t)+\mathcal{I}(t)\mathcal{Q}\rho(t_0),
\end{equation}
where the superoperators are defined as
\begin{eqnarray*}
\mathcal{K}(t)=\mathcal{P}\mathcal{L}(t)[1-\Sigma(t)]^{-1}\mathcal{P},\\
\mathcal{I}(t)=\mathcal{P}\mathcal{L}(t)[1-\Sigma(t)]^{-1}\mathcal{G}(t,t_0)\mathcal{Q},\\
\Sigma(t)=\int_{t_0}^t\mathcal{G}(t,s)\mathcal{Q L}(s)\mathcal{P}G(t,s)ds,\\
G(t,s)=\exp_+\left[-\int_{s}^t \mathcal{L}(s')ds'\right], \\
\mathcal{G}(t,s)=\exp_-\left[\int_{s}^t \mathcal{QL}(s')ds'\right].
\end{eqnarray*}
In the above expressions $\mathcal{L}(t)A=-i[H(t),A]$ is the Liouville superoperator and $H(t)$ is the system Hamiltonian,
 $\exp_{\mp}$ is the chronological (antichronological) exponent, $\mathcal{Q}=1-\mathcal{P}$ and $\mathcal{P}$ is some projection operator (we set $\hbar=k_B=1$).

Eq.~(\ref{TCL}) is exact, but the calculation of the right-hand side is associated with  difficulties. The usual approach to Eq.~(\ref{TCL}) is a perturbation expansion of the right-hand side and the consideration of a few perturbation orders. Often the second order perturbation expansion is used and the master equation has the following form
\begin{eqnarray}
\mathcal{P}\dot{\rho}=\mathcal{PL}(t)\left\lbrace [1+\int_{t_0}^t ds\mathcal{QL}(s)]\mathcal{P}\rho(t)\right. \label{TCL2}\\
+\left.[1+\int_{t_0}^t ds\mathcal{QL}(s)]\mathcal{Q}\rho(t_0)\right\rbrace. \nonumber
\end{eqnarray}
This equation reproduces the majority of Markovian and non-Markovian dynamical master equations under different assumptions \cite{toqs}.

\section{Projection Operators}
Clearly, the general form of Eq.~(\ref{TCL}) and (\ref{TCL2}) does not depend on the concrete form of the projection operator $\mathcal{P},$ and here one has a lot of freedom to choose the most suitable one. Traditionally, the projection operator for open quantum systems is chosen in the form 
\begin{equation}\label{trad}
\mathcal{P}A=\mathrm{Tr}_E(A)\otimes\frac{\exp[-\beta H_E]}{\mathrm{Tr}\exp[-\beta H_E]},
\end{equation}
where the partial trace takes over the environment degrees of freedom. In the above equation $H_E$ is the free Hamiltonian of the environment and $\beta$ is the inverse temperature.  Application of the traditional projection operator (\ref{trad}) to the TCL master equation ~(\ref{TCL2}) 
is in general equivalent to the well-known Born approximation, and physically means the absence of any dynamics of the environment.

Recently, several attempts to optimise the perturbation theory by using  other projectors were made \cite{CRL, MY1}. Particularly, the action of the correlated projection operator \cite{CRL} takes into consideration additional degrees of freedom of the environment. But every single additional degree of freedom leads to the appearance of an additional equation in the system (\ref{TCL}), which restricts the applicability of the method. 
Another type of a projector is the Kawasaki-Gunton projection operator \cite{MY1,KAWASAKI}, which is based on ideas of non-equilibrium thermodynamics. This projector leads in general to a non-linear system of equations, and of course, such a system is not easy to study.

In this letter we suggest the maximal possible simplification of Eq.~(\ref{TCL}).  As in the traditional approach based on the projection operator (\ref{trad}) we ignore any changes in the environment. The basic idea is to study every relevant degree of freedom of an open quantum system separately. A projection operator which extracts only one degree of freedom 
has the from
\begin{equation}\label{proj}
\mathcal{P}_{ij}A=\mathrm{Tr}\left\lbrace (E_{ij}^\dagger\otimes I) A\right\rbrace E_{ij}\otimes\frac{\exp[-\beta H_E]}{\mathrm{Tr}\exp[-\beta H_E]},
\end{equation} 
where $I$ is identity matrix acting in the Hilbert space of the environment and $E_{ij}$ is a matrix with unit in the intersection of the $i$th row and $j$th column and 0 elsewhere acting in the Hilbert space of the open system.  
 
Clearly, Eq.~(\ref{TCL}) with the projection operator (\ref{proj}) reduces to a single linear differential equation, which can be easy solved in quadratures, namely
\begin{eqnarray}
\mathcal{P}_{ij}\rho(t)=\exp\left[\int_{t_0}^t\mathcal{K}_{ij}(s)ds \right]\mathcal{P}_{ij}\rho(t_0) \label{sol}\\
+\int_{t_0}^tds\exp\left[\int_{s}^t\mathcal{K}_{ij}(s')ds' \right]\mathcal{I}(s)\mathcal{Q}_{ij}\rho(t_0)ds,\nonumber
\end{eqnarray}
where the subscripts denote the projection operator, used in the definition of the superoperators, and $t_0$ is the initial moment of time.

The expression (\ref{sol}) is exact and, in principle, can be calculated with any accuracy. By changing the projection operator in Eq.~(\ref{sol}) one can find any element of the density operator. The hermitian character of the density operator and also the normalisation conditions allow to decrease the number of necessary elements to completely characterise an open system. The number of the elements  needed is equal to $(n-1)(n+2)/2,$ which consists of $(n-1)$ diagonal elements of the density operator and $n(n-1)/2$ of non-diagonal elements below (above) the diagonal.

\section{Iterative procedure for recovering of the reduced density operator}
The exact calculation of the superoperators in Eq.~(\ref{sol}) is associated with significant difficulties. Thus, one has to restrict oneself to some perturbation expansion up to appropriate order. In this section we suggest an iterative procedure which allows to recover the exact dynamics using the appropriate perturbation expansion.

The recursive procedure is constructed as follow. The perturbation expansion of Eq. (\ref{sol}) with the projector (\ref{proj}) {approximates} the exact dynamics on some time interval, which is usually defined by the strength of the interaction between the components of the system, i.e. if the maximal interacting constant is $\lambda$ than the expansion is valid for $\lambda t\ll 1$. Let the initial time be denoted by $t_0$ and $t_f$ be the final time, where the perturbation expansion is still applicable. Beyond the interval  $[t_0,t_f]$ the omitted degrees of freedom start to significantly affect on the dynamics.
Take some point $t_1$ in the valid interval and consider the values of Eq.~(\ref{sol}) in this point as new initial conditions one can again using Eq.~(\ref{sol}) extend the solution to the new time interval $[t_1,t_2].$ Continuing this procedure one can iteratively reproduce the dynamics for any time interval $[t_0,t].$ In other words, the iteration procedure is
\begin{eqnarray}
\mathcal{P}_{ij}\rho(t_k)=\exp\left[\int_{t_{k-1}}^{t_k}\mathcal{K}_{ij}(s)ds \right]\mathcal{P}_{ij}\rho(t_{k-1}) \label{sol2}\\
+\int_{t_{k-1}}^{t_k}ds\exp\left[\int_{s}^{t_k}\mathcal{K}_{ij}(s')ds' \right]\mathcal{I}(s)\mathcal{Q}_{ij}\rho(t_{k-1})ds,\nonumber
\end{eqnarray}
where the superoperators are defined perturbatively. For instance, up to the second order we have
\begin{eqnarray}
\mathcal{K}_{ij}(t)=\mathcal{P}_{ij}\mathcal{L}(t)\left\lbrace [1+\int_{t_0}^t ds\mathcal{Q}_{ij}\mathcal{L}(s)]\mathcal{P}_{ij}\right\rbrace,\label{K2}\\
\mathcal{I}_{ij}(t)=\mathcal{P}_{ij}\mathcal{L}(t)\left\lbrace [1+\int_{t_0}^t ds\mathcal{Q}_{ij}\mathcal{L}(s)]\mathcal{Q}_{ij}\right\rbrace. \label{I2}
\end{eqnarray}

{The perturbation expansions (\ref{K2})-(\ref{I2}) reproduce the exact expressions with the order $o(\lambda^2 t^2)$, where $\lambda >||\mathcal{L}||.$ Thus the iterative scheme (\ref{sol2}) has an error $\sim o(\lambda^2 (t_i-t_{i-1})^2)$ in general.} 

Notice, that we  ignore any changes in the environment. Thus, to effectively apply the iteration procedure one has to define $(n-1)(n+2)/2$ matrix elements, characterising the open quantum system. With this knowledge one can calculate $\mathcal{Q}_{ij}\rho(t_{k-1})$ and continue the iterations.
In this case the suggested iterative scheme allows to reproduce the solution of traditional forms of Markovian and non-Markovian master equation, which follows from (\ref{TCL2}) for any arbitrary time-interval and with arbitrary precision.  {Also notice that the inhomogeneity $\mathcal{Q}_{ij}\rho(t_{i-1})$ cannot be neglected in general.}
Below we show two examples of the application of the above theory.

\section{Non-Markovian relaxation of a qubit}
The simplest example of a non-Markovian process is relaxation of a qubit in a thermal environment.
The Hamiltonian of the system in the interaction picture is 
\begin{equation}
H=\sigma_-B^\dagger(t)+\sigma_+ B(t),
\end{equation}
where $\sigma_\pm $ are the Pauli matrices, and $B(t)=\sum_ke^{i(\omega_0-\omega_k)t}b_k$, $b_k$ is the annihilation operator of the $k$th oscillator in the bath, $\omega_0$ is the transition frequency of the qubit and $\omega_k$ is the frequency of the $k$th oscillator in the bath. The dimension of the Hilbert space of the qubit is equal to 2. Thus,
for recovering the density operator with the help of the above procedure one has to find $(n-1)(n+2)/2=2$  matrix elements. The corresponding projector operators (\ref{proj}) have the following form
\begin{eqnarray}
\mathcal{P}_{11} A&=&\mathrm{Tr}(A \sigma_+\sigma_-)\sigma_+\sigma_-\otimes\rho_B,\\
\mathcal{P}_{21} A&=&\mathrm{Tr}(A \sigma_+)\sigma_-\otimes\rho_B,
\end{eqnarray}
where $\rho_B=\exp[-\beta\sum_k \omega_k b_k^\dagger b_k]/\mathrm{Tr}\exp[-\beta\sum_k \omega_k b_k^\dagger b_k].$

Substituting the above projectors for Eq.~(\ref{TCL2}) leads to the following equations for the factorised system-bath initial conditions
\begin{eqnarray}
\dot{\rho}_{11}&=&-f_+(t)\rho_{11}(t)+f_-(t)\rho_{22}(t_0), \label{EQ1}\\
\dot{\rho}_{21}&=&-g(t)\rho_{21}(t).\label{EQ2}
\end{eqnarray}
In the above equations $f_\pm(t)=\int_0^\infty d\omega J(\omega)(\mathrm{coth}(\beta\omega/2)\pm 1)\sin[(\omega-\omega_0)t]/(\omega-\omega_0),$ and $g(t)=\int_0^\infty d\omega\left( 1-\exp[i(\omega-\omega_0)t]\right) J(\omega)\mathrm{coth}(\beta\omega/2)/(\omega-\omega_0),$ where $J(\omega)$ is the spectral density of the bath. It is clear that the solution of Eqs.~(\ref{EQ1})-(\ref{EQ2}) can be found in quadratures.

It is interesting to compare the above result with the standard results following from the TCL master equation (\ref{TCL2}) with the projector (\ref{trad}). The traditional  equation for $\rho_{21}$ coincides with (\ref{EQ2}), but the result for $\rho_{11}$ differs from Eq.~(\ref{EQ1}). To reproduce the result following from the standard TCL equation we have to replace $\rho_{22}(t_0)$ by $\rho_{22}(t)$ in the right-hand side of Eq.~(\ref{EQ1}). Actually, it is the general rule that the master equation with the projector (\ref{proj}) has the same structure as the traditional master equation with the projector (\ref{trad}), but all ``non-relevant'' variables are replaced by its initial values. Also, the restricted superoperator leads to break the trace preserving of the the dynamical map. It can be checked that for the considered model $\mathrm{Tr}\dot{\rho}=f_+(t)(\rho_{11}(t_0)-\rho_{11}(t))+f_-(t)(\rho_{22}(t_0)-\rho_{22}(t))$ is not equal to zero at any time $t>t_0.$ {The order of this effect is $o(l(t-t_0)^2),$ where $l>\max(|f_-|,|f_+|)$.} Thus, one can make the value of the trace arbitrary small, taking the time interval $[t,t_0]$ small enough and apply the iteration procedure, described in the previous section.

Replacing $\rho_{22}(t_0)$ in the right hand side of Eq.~(\ref{EQ1}) by $1-\rho_{11}(t_0)$ and applying the iterative procedure from the previous section, which for this model has the form
\begin{eqnarray}
\rho_{11}(t_k)=\exp\left[-\int_{t_{k-1}}^{t_k}f_+(s)ds\right]\rho_{11}(t_{k-1})\\
+\int_{t_{k-1}}^{t_k}ds\exp\left[-\int_{s_1}^{t_k}f_+(s_1)ds_1\right] f_-(s)(1-\rho_{11}(t_{k-1})),\nonumber
\end{eqnarray}
one can recover the results of the standard non-Markovian master equation. 

The dynamics of the excited state of the qubit for the Ohmic spectral density $J(\omega)=\lambda\omega\exp[-\omega/\Omega]$ is presented in Fig. 1. One can see that the iterative scheme (\ref{sol2}) reproduces the solution of the traditional master equation for the considered system quite well. The time step for the iterations was $\lambda t=0.05$ and the absolute error of the approximation does not exceed $0.01.$ {In the same figure we draw the Markovian limit, which  corresponds to the parameters $f_\pm(+\infty).$ One can see that in the Markovian limit the approach also works well.}


\section{Two interacting qubits in a common thermal bath}
The example in the previous section is very simple and one can build the analytical solution of the traditional form of the master equation in quadratures. In this section we consider a more complicated example. The Hamiltonian for the system is written as 
\begin{equation}\label{ham2}
H=H_0+H_{12}+H_B+H_{int},
\end{equation}
where $H_0= \omega_0\sum_i\sigma_{z}^i$ is the free qubits Hamiltonian, $H_B=\sum_j \omega_j b^\dagger_j b_j$ is the free Hamiltonian of the bath, $H_{12}=V(\sigma_+^1\sigma_-^2+\sigma_+^2\sigma_-^1)$ is the Hamiltonian of the qubit interactions and $H_{int}=\sum_i\sum_j g_{j} b_j \sigma_+^i\alpha_{i}+\mathrm{h.c.}$ is the qubits-bath interaction Hamiltonian, $\sigma^i$ is the Pauli matrices for the $i$th qubit, $b_j$ is the annihilation operator of the $j$th oscillator in the bath, $ \omega_0$  and $\omega_j$ are the transition frequency of the qubits and the $j$th oscillator in the bath, correspondingly, $V$ is the constant of dipole-dipole interaction, $g_{j}$ is the coupling constant of the qubit and $j$th oscillator in the bath, $\alpha_i$  are the geometrical factors that mark the position of the $i$th qubit. 

\begin{figure}
\includegraphics[scale=0.6]{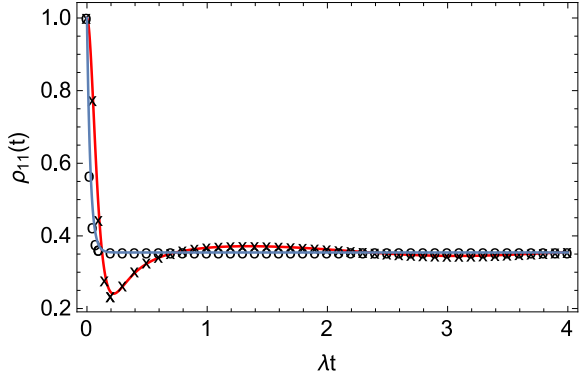}
\caption{(colour online) The evolution of the exicited state $\rho_{11}(t)$ of the single qubit in the thermal bath. Red curve is the  solution of Eq.(\ref{TCL2}) with the projector (\ref{trad}), x-signs iterative solution of Eq. (\ref{EQ1}) with the time step $\lambda t= 0.05$. {Blue curve and circles are the Markovian limit for projectors (\ref{trad}) and (\ref{proj}) respectively.} Parameters in the system are $\Omega=10\lambda,\, \omega_0=2\lambda, \, \beta=0.3$  } \label{result}
\end{figure}

First we transform the Hamiltonian (\ref{ham2}) to the interaction picture
\begin{eqnarray}
H_I(t)=e^{i(H_0+H_{12}+H_B)t}H_{int}e^{-i(H_0+H_{12}+H_B)t}\label{hami}\\
=(P_+^1R(\alpha_2,-\alpha_1)+P_-^1R(\alpha_2,\alpha_1))\sigma_+^2B(t)\nonumber\\
+(P_+^2R(\alpha_1,-\alpha_2)+P_-^2R(\alpha_1,\alpha_2))\sigma_+^1B(t)+\mathrm{h.c.}\nonumber\\
=K(t)B(t)+K^\dagger(t)B^\dagger(t),\nonumber
\end{eqnarray}
where $P^i_+=\sigma^i_+\sigma^i_-,$ $P^i_-=\sigma^i_-\sigma^i_+$ and $R(\alpha,\beta)=\alpha\cos(tV)+i\beta\sin(tV),$  $K(t)=\left(P_+^1R(\alpha_2,-\alpha_1)+P_-^1R(\alpha_2,\alpha_1)\right)\sigma_+^2+(1\leftrightarrow 2)$, $B(t)$ is the same as for the previous model.

\begin{figure}
\includegraphics[scale=0.6]{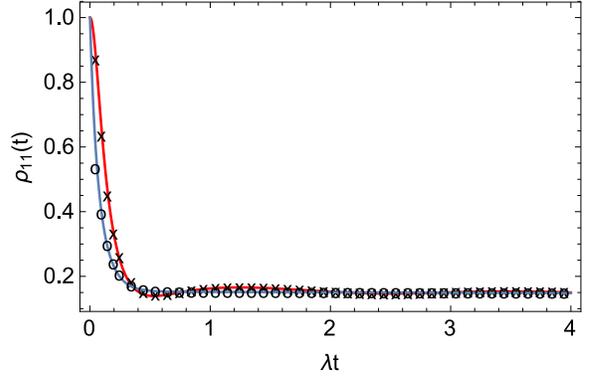}
\caption{(colour online) The evolution of the collective exited state $\rho_{11}(t)$ of two interacting qubits in the common thermal bath. Red curve is the  solution of Eq.(\ref{ME2}) with traditional projection operator (\ref{trad}), x-signs the iterative procedure with the help of Eq. (\ref{sol2}) with the time step $\lambda t= 0.05$. {Blue curve and circles are the Markovian limit for projectors (\ref{trad}) and (\ref{proj}) respectively.} Parameters in the system are $V=0.6\lambda, \, \Omega=10\lambda,\, \omega_0=2\lambda, \, \beta=0.3, \, \alpha_1=0.4+0.3i, \,\alpha_2=0.5+0.2i$  } \label{result}
\end{figure}

The master equation (\ref{TCL2}) for the factorised initial conditions can be written as
\begin{eqnarray}
\dot{\mathcal{P}\rho}(t)\!\!\!&=\!\!\!\int^t_0 dt_1\mathcal{P}\left\{[K^\dagger(t_1)\mathcal{P}\rho(t)K(t)-K(t)K^\dagger(t_1)\mathcal{P}\rho(t)]L\right.\nonumber\\
+&\!\!\!\!\!\!\!\!\!\!\!\!\!\!\!\left.[K(t_1)\mathcal{P}\rho(t)K^\dagger(t)-K^\dagger(t)K(t_1)\mathcal{P}\rho(t)]N+\mathrm{h.c}\right\},\label{ME2}\!\!\!\!\!\!\!\
\end{eqnarray}
where $F=\int_0^\infty d\omega J(\omega)(\coth(\beta\omega/2)+1)/2e^{i(\omega_0-\omega)(t-t_1)}$ and $R=\int_0^\infty d\omega J(\omega)(\coth(\beta\omega/2)-1)/2e^{-i(\omega_0-\omega)(t-t_1)},$ and we assumed that the bath stays in the thermal equilibrium for all time. The projection operator $\mathcal{P}$ in the above equation is either (\ref{trad}) or (\ref{proj}).

One can see that the investigation of the  master equation (\ref{ME2}) with the projection (\ref{trad}) is not a trivial problem. Even numerical solution of the equation is quite a tricky and consist of  the evaluation of the multidimensional integrals in every step. The application of the projection operators (\ref{proj}) leads to a uncoupled system of equations, which has the same form as the system following from Eq.~(\ref{ME2}). The only difference is that all "irrelevant"  matrix elements for the concrete projection operator in the form (\ref{proj}) are replaced by its initial values.

As it was mentioned above, to completely describe the open systems one has to solve $(n+2)(n-1)/2=9$ uncoupled equations. By using the iteration scheme (\ref{sol2}) one can reproduce the dynamics of the open system. The result for the spectral density $J(\omega)=\lambda\omega e^{-\omega/\Omega}$ for the population of the collective excited state is shown in Fig.~2. {The Markovian dynamics follows from (\ref{ME2}) changing the upper limit of integration to  $+\infty$.} One can see that for the time step $\lambda t=0.05$ the iteration procedure  (\ref{sol2}) gives very accurate result. 

{\section{Generalization}
Above we discussed the method only in application to a bosonic bath. In this paragraph we want to stress the possibility of the method for describing other types of open quantum systems. Of particular interest is the case of environment with specific spectral properties \cite{BUDINI,MY,Breuer}. In this case it is necessary to take into account the evolution of some environmental degrees of freedom. This can be done by generalisation of the projection operator (\ref{trad}) in the following way. Let the state of the bath be characterised by some set of $r$ orthonormal vectors $|a_i\rangle, i=1, ... , r.$ The projection operators which allow to study such a system have the form 
\begin{equation}\label{CP}
\mathcal{P}_i=\mathrm{Tr}_E(|a_i\rangle\langle a_i|\rho)\otimes |a_i\rangle\langle a_i|.
\end{equation}
By substituting this projector into (\ref{TCL}) leads to  $n^2 r$ differential equations in any order of perturbation expansion. The reduced density operator is defined by $\rho_S(t)=\sum_i \mathcal{P}_i\rho(t).$ 

The above theory also can be modified for this case. To extract only one degree of freedom from the density operator and take into account  bath degrees of freedom we introduce the projection operator
\begin{equation}\label{CP2}
\mathcal{P}_i^{kl}A=\mathrm{Tr}(E_{kl}^\dagger\otimes |a_i\rangle\langle a_i| A)E^{kl}\otimes |a_i\rangle\langle a_i|.
\end{equation}
One has to consider only $r(n^2+n)/2$ independent linear equations due to identity $\mathcal{P}_i^{kl}\rho=\mathcal{P}_i^{lk}\rho$. The dynamics of the $kl$th  degree of freedom of the open system is equal to $\rho^{kl}_S(t)=\sum_i \mathcal{P}_i^{kl}\rho(t).$ Notice that in this case the number of equations needed to describe the system decreases significantly.
Applying the iterative scheme one can reproduce  the solution of the master equation with the projector (\ref{CP}) of corresponding order. Thus the general idea of this letter can be easily modified to describe other open quantum systems.

}

\section{Conclusions}
The general form of the TCL master equation does not depend on the concrete form of the projection operator. This fact has been used to sufficiently simplify the study of the dynamics of open quantum systems by choosing the projection operator which extracts only one degree of freedom of the density operator. Substituting the simplest form of the projection operator for the TCL master equation always leads to a single linear differential  equation, which can be  solved in quadratures in any perturbation order.  Thus, one can efficiently study every degree of freedom of an open quantum system separately, and due to hermicity and trace preservation of the density operator the number of the necessary degrees of freedom to completely describe a system can be reduced to $(n+2)(n-1)/2,$ where $n$ is the dimension of the system Hilbert space. In general, the accuracy of the TCL expansion depends only on the perturbation order, but not on the concrete form of a projection operator. Thus, the solution of the TCL master equation with the simplest projection operator (\ref{sol}) reproduces the dynamics with the same accuracy as the traditional form of the master equations of the corresponding order.

Additional assumptions, such as constant state of the environment during the evolution of an open quantum system and week system-environment interaction, allow to truncate the Hilbert space of the total system and to consider only the second order perturbation expansion. In such a case the traditional form of the TCL master equation gives reasonably good results, in particular these results consist of the Lindblad form of the Markovian master equation. We have suggested the iteration scheme (\ref{sol2}), which allows to recover the results following from the  traditional TCL master equations and shown with concrete examples that the suggested scheme allows to reproduce the  traditional second order master equation with a good precision.

\acknowledgments
This work is based upon research supported by the South African
Research Chair Initiative of the Department of Science and
Technology and National Research Foundation.


\begin{thebibliography}{0}

\bibitem{PS}  \Name{Jeske J., Ing D. J., Plenio M. B.,  Huelga S. F. and  Cole J.H.} \REVIEW{J.  Chem. Phys.}{142}{2015}{064104}.

\bibitem{transfer}  \Name{Sinayskiy I., Marais A.,  Petruccione F. and Ekert A.}  \REVIEW{Phys. Rev. Let.}{108}{2012}{020602}.

\bibitem{LF}  \Name{Meier  C. and  Tannor D. J.}  \REVIEW{J. Chem. Phys.}{111}{1999}{3365}.

\bibitem{MY1} \Name{Semin V.  and Petruccione F.}  \REVIEW{Phys. Rev. A}{90}{2014}{052112}.

\bibitem{KOSSAK} \Name{Chru\'{s}ci\'{n}ski D. and  Kossakowski A.}  \REVIEW{Phys. Rev. Lett. }{111}{2013}{050402}.

\bibitem{KOCH} \Name{ Li A.C.Y., Petruccione F.  and Koch  J. } \REVIEW{Sci. Rep.}{4}{2014}{4887}.

\bibitem{toqs} \Name{Breuer H.-P.  and Petruccione F. }, \Book{The Theory of Open Quantum Systems} \Publ{Oxford University Press,Oxford} \Year{2002}.

\bibitem{CAO} \Name{Moix J.M. and  Cao J.},  \REVIEW{J. Chem. Phys}{139}{2013}{134106}.

\bibitem{BUDINI} \Name{Budini A.A.}  \REVIEW{Phys. Rev. A}{88}{2013}{012124}.

\bibitem{BREUER} \Name{Wissmann S. , Karlsson A., Laine E.-M.,  Piilo J. and Breuer H.-P. }   \REVIEW{Phys. Rev. A}{86}{2012}{062108}.

\bibitem{CRL} \Name{Fischer J. and Breuer H.-P. }  \REVIEW{Phys. Rev. A}{76}{2007}{052119}.

\bibitem{KAWASAKI} \Name{Kawasaki K. and Gunton  J.D.} \REVIEW{Phys. Rev. A}{8}{1973}{2048}. 

{\bibitem{BUDINI} \Name{Budini A.}\REVIEW{Phys. Rev. E}{89}{2014}{012147}.

\bibitem{MY} \Name{Semin V., Sinayskiy I. and Petruccione}\REVIEW{Phys.Rev. A}{86}{2012}{062114}.

\bibitem{Breuer} \Name{Breuer H.P.} \REVIEW{Phys. Rev. A}{75}{2007}{022103}

}

\end{thebibliography}
\end{document}